\documentclass[prb,twocolumn,showpacs,preprintnumbers,amsmath,amssymb,10pt]{revtex4-1}

\usepackage{color}
\definecolor{blau}{rgb}{0,0,1}

\usepackage[T1]{fontenc}
\usepackage[pdftex]{graphicx}
\usepackage{dcolumn}
\usepackage{bm}

\usepackage{mathcomp}

\DeclareGraphicsExtensions{.pdf}   
\usepackage[justification=raggedright,singlelinecheck=false]{caption}

\begin{document}

\title{Band structure calculations of CuAlO$_2$, CuGaO$_2$, CuInO$_2$ and CuCrO$_2$ by screened exchange}
\author{Roland Gillen}\email{rg403@cam.ac.uk}
\author{John Robertson}
\affiliation{Department of Engineering, University of Cambridge, Cambridge CB3 0FA, United Kingdom}

\date{\today}

\begin{abstract}
We report density functional theory (DFT) band structure calculations on the transparent conducting oxides CuAlO$_2$, CuGaO$_2$, CuInO$_2$ and CuCrO$_2$. The use of the hybrid functional sX-LDA leads to considerably improved electronic properties compared to standard local density approximation (LDA) and generalized gradient approximation (GGA) approaches. We show that the resulting electronic band gaps compare well with experimental values and previous quasiparticle calculations and show the correct trends with respect to the atomic number of the cation (Al, Ga, In). The resulting energetic depths of Cu $d$ and O $p$ levels and the valence band widths are considerable improvements compared to LDA and GGA and in good agreement with available x-ray photoelectron spectroscopy (XPS) data. Lastly, we show the calculated imaginary part of the dielectric function for all four systems.
\end{abstract}


\maketitle

\section{Introduction}
"Invisible electronic devices" which allow for novel flat panel systems and improved solar cells are an interesting new field in optoelectronics. Such systems require transparent or nearly transparent materials with band gaps in the range of 3\,eV or above and good $n$- and $p$-type dopability. Transparent conductive oxides (TCOs) are promising candidate materials and have excited intensive research during the previous decade for this reason. Good $n$-type dopability is achievable in ZnO, In$_2$O$_3$\cite{hosono-2006,nomura-2006} and SnO$_2$\cite{robertson-2008}, but $p$-type doping still is difficult. 
The tendency of oxides to form non-bonding O 2p states with a high effective mass at the valence band maximum is problematic, as a low effective mass is needed ionizable shallow acceptor states\cite{kawazoe-1997}. 
In 1997, Kawazoe \emph{et al.}\cite{kawazoe-1997} reported $p$-type conductivity in combination with transparency in CuAlO$_2$ and similar properties were discovered for other members of the CuMO$_2$ (M=Ga,In,Cr etc.) group. These materials possess inherent advantages over Cu$_2$O: (i) Their optical band gaps are 3\,eV and above, (ii) the additional cations M are likely to stabilize the oxygen atoms in the compound and contribute to dopability\cite{robertson-2011-doping}.


Despite the many scientific reports on CuMO$_2$, details of the band structure and the electronic properties, such as hole conductivity mechanism and the abundance of compensating defects in these materials are still unclear. 
Here, \emph{ab initio} calculations are valuable. 
However, the common density functional theory (DFT) calculations within the local density approximation (LDA) or the generalized gradient approximation (GGA) lack the derivative discontinuity with respect to fractional charges\cite{cohen-2008} and thus suffer from spurious self-interaction. This self-interaction promotes artificial delocalization of electron states and causes occupied states to be placed too high in energy, lowering the size of the predicted band gaps. This limits the predictive abilities of these approximations. It is thus necessary to go beyond LDA and GGA for the theoretical investigation of the electronic properties.

In this report, we present calculations of the electronic properties of four delafossite TCOs, CuAlO$_2$, CuGaO$_2$, CuInO$_2$ and CuCrO$_2$ employing the screened-exchange-LDA (sX-LDA) hybrid functional\cite{byklein-SX,Seidl-SX}. The inclusion of exact Hartree-Fock exchange compensates for the self-interaction error and has a beneficial effect on the predicted electronic properties. (i) We show that sX-LDA noticeably improves on the band gap energies compared to LDA/GGA for all studied TCOs and compare our values with those from other methods. (ii) The experimentally observed trend of the band gaps in the sequence CuAlO$_2$-CuGaO$_2$-CuInO$_2$ is reproduced by our sX-LDA calculations. (iii) The predicted valence band widths (with possibly the exception of CuCrO$_2$) and (iv) the depths of the Cu d levels are close to experimental values. Further, we provide the calculated imaginary part of the dielectric function for all studied materials. We identify a strong renormalization of the d state energy to be the main factor in our calculations.

\section{Method}
There are various methods to improve the band gaps. A widespread and computationally efficient approach is LDA + U\cite{anisimov-1991-lda+u}, where a empirical on-site Coulomb energy is added to selected orbitals. The energetical down/upshift of the corresponding bands compared to pure LDA results in improved band gaps. Green's function approaches such as GW\cite{Hedin-GW} and its approximations\cite{GW-method} explicitly address the many-body problem and treat electrons as quasiparticles. The aim is then to calculate the electron self-energy in terms of the single particle Green's function G and a dynamically screened Coulomb interaction W. While the resulting quasiparticle energies lead to accurate band gaps, the frequency/energy-dependent screening in W makes the method computationally expensive.

An alternative is to include a fraction of screened Hartree-Fock exchange in otherwise purely density-dependent exchange-correlation functionals. The idea of such hybrid functionals is a best-of-both-worlds approach, to solve the band gap problem by combining the overestimation of band gaps from Hartree-Fock with the underestimation from LDA and GGA. HSE\cite{HSE-03} is such a functional, which recently gained large popularity. It yields good results for small- and medium band gap semiconductors. 

In this report, we use the hybrid functional sX-LDA. 
The sX-LDA method was proposed by Bylander and Kleinman\cite{byklein-SX} as a modification of the local density approximation on empirical grounds. Seidl \emph{et al.} later showed that the method can be derived from a generalized Kohn-Sham scheme. The idea is to 
split the exchange-correlation potential in (i) an orbital-dependent short-range term, which can be treated exactly, and (ii) am explicit density functional term, but at the same time to recover the accurate LDA behavior for homogenous electron gases. The exchange-correlation potential can then be written as 
\begin{equation}
V_{XC}^{sX-LDA} = V_x^{sX} + V_{xc}^{LDA} - V_x^{sX,local}\label{eq:sX}
\end{equation}
describes the exchange interaction by a statically screened exact-exchange term
\begin{equation}
V_x^{sX} = \sum_{i,j}\int
d\mathbf{r}\frac{\psi_{i}^*(\mathbf{r})e^{k_{s}|\mathbf{r}-\mathbf{r}'|}\psi_{j}
(\mathbf{r})}{|\mathbf{r}-\mathbf{r}'|}
\end{equation}
, usually with the Thomas-Fermi wave vector k$_s$=k$_{TF}$ as inverse screening length. 
In our calculations, we used a value of k$_s$=0.5 bohr$^{-1}$, which works well for transition metal compounds. 
The correlation energy in this functional is purely LDA. 

The sX-LDA method has recently been implemented\cite{clark-sx} within the plane-wave
basis set in the DFT package CASTEP\cite{castep1} and was shown to yield band gaps that can compete in accuracy with those from G$_0$W$_0$.
The potential of the ions was modelled by OPIUM-generated normconserving pseudopotentials and the electrons where represented by plane-waves with a cut-off energy of 800\,eV for all structures. We averaged the integration over reciprocal space using a grid of 4x4x4 evenly distributed k-points in the Brillouin zone. The PW91 and sX-LDA band structures were obtained by ground-state minimization and subsequent band structure
calculation, with the exception of CuCrO$_2$, where we had to obtain the sX-LDA band structure by perturbation of the spin-polarized PW91 ground state for speed reasons.

The CuMO$_2$ TCOs crystallize in the delafossite structure, consisting of planes of O atoms caged in tetrahedra made of M anions and Cu atoms. The planes are connected by dumbbell-like O-Cu-O bridges. 
Depending on whether the layer stacking is AB or ABC, the structure is of the
2H-type with hexagonal $P6_3/mmm$ symmetry, or of the 3R-type with rhombohedral
$R\bar{3}m$ symmetry. The energy-difference between both structures is low. We modelled our studied TCOs by a rhombohedral unit cell 
with the lattice vectors $a_1=(\frac{a}{2},-2\sqrt{3}a,\frac{c}{3})$,
$a_2=(-\frac{a}{2},-2\sqrt{3}a,\frac{c}{3})$, $a_3=(0,2\sqrt{3}a,\frac{c}{3})$ 
atoms at the Wyckoff positions 
.
A test wise optimization of the cell parameters of CuAlO$_2$ showed a very small change of 0.7\% compared to the lattice
parameters and we decided to use the experimental values for all our
calculations in order to be consistent with calculations from other groups. On
this basis, we relaxed the atomic positions until the residual forces were
smaller than $10^{-3}$\,eV/\AA. We further used a grid of 15x15x15 equidistant points in the Brillouin zone for the optical calculations and a 9x9x9 grid for the calculations of the density of states. 

\section{Results and Discussion}
\subsection{Band structures of CuAlO$_2$,CuGaO$_2$ and CuInO$_2$}
\begin{figure}
\begin{minipage}{0.488\columnwidth}
\centering
\includegraphics*[width=\columnwidth]{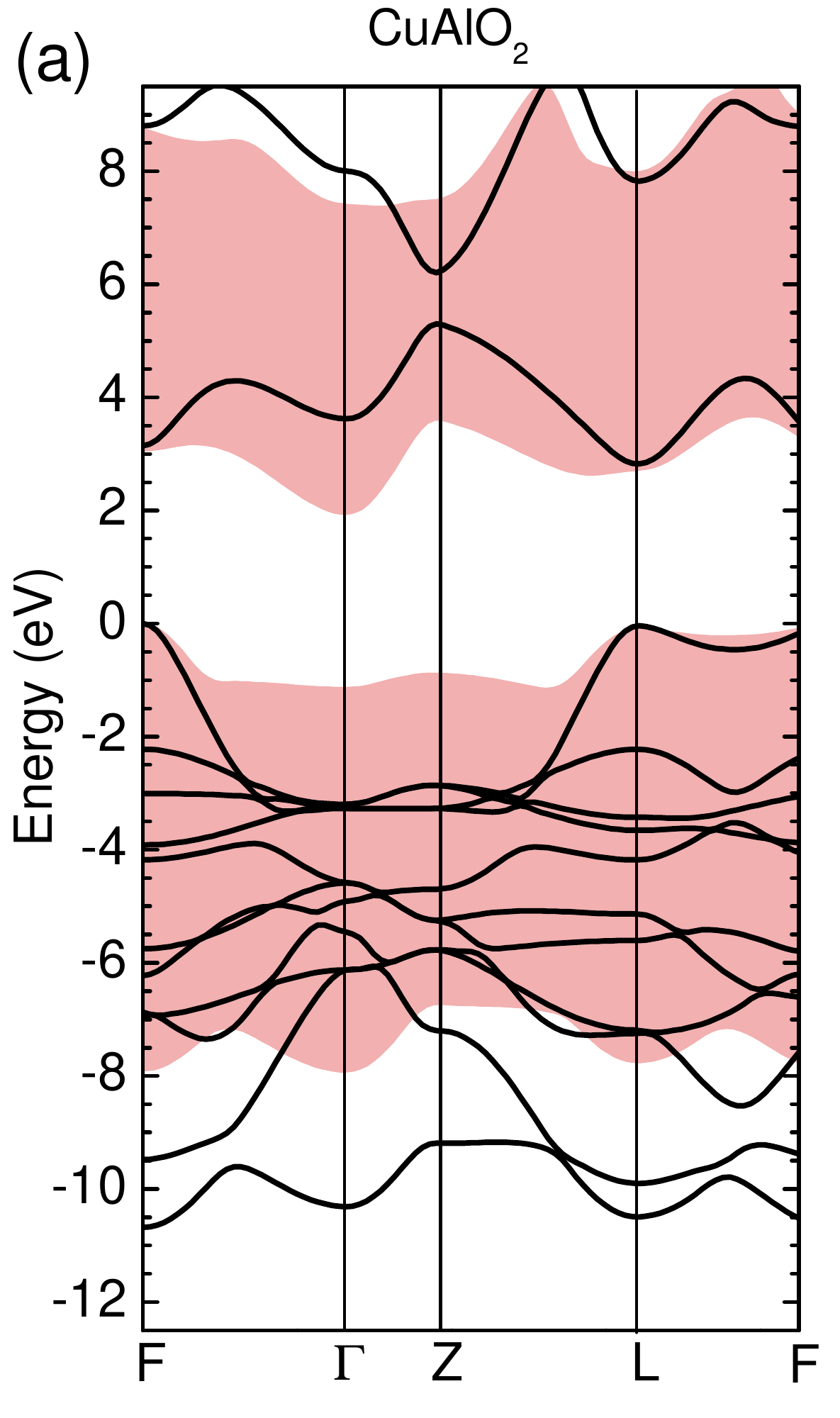}
\end{minipage}
\space
\begin{minipage}{0.488\columnwidth}
\centering
\includegraphics*[width=\columnwidth]{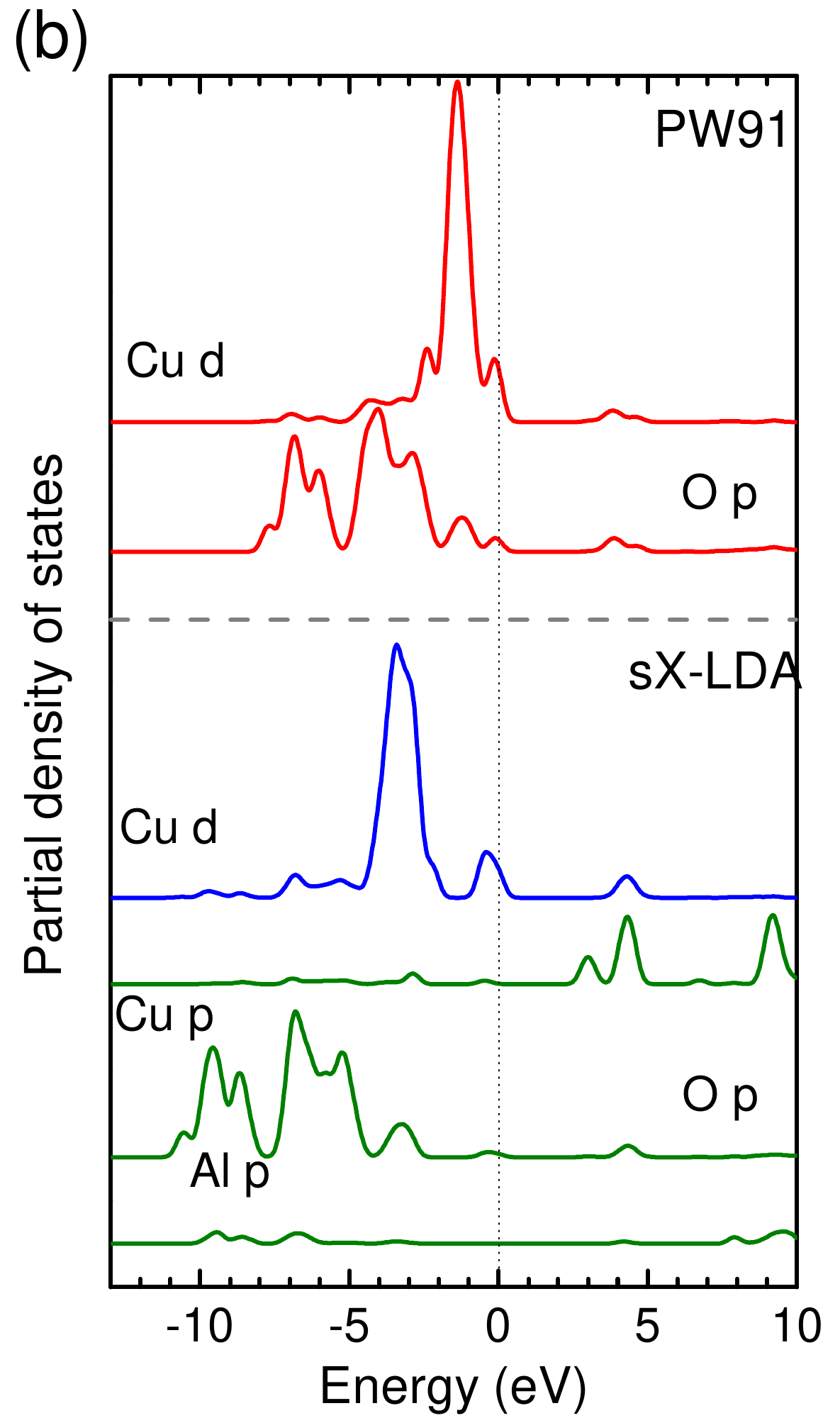}
\end{minipage}
\caption{\label{fig:CuAlO2-sXLDA} (Color online) (a) Electronic band structures of CuAlO$_2$ from sX-LDA (black solid lines) and GGA (shaded area) calculations. (b) Corresponding partial density of states of the dominant angular momentum channels.}
\end{figure}
We start our discussion with CuAlO$_2$, certainly the most prominent delafossite material.
Available experimental values for the optical band gaps of CuAlO$_2$ are quite disperse and range between 2.9\cite{yu-2007} and 3.9\cite{banerjee-2005} eV, where most studies point to a gap of 3.5-3.6\,eV\cite{kawazoe-1997,yanagi-2000,tate-2009}. 
Theoretical studies point towards the existence of an indirect fundamental band gap that optically inactive due to symmetry reasons\cite{nie-2002}. Indeed, evidence for such an indirect band gap was found experimentally, although the band gap size is subject to debate. Various experimental studies\cite{benko-1984,yanagi-2000} on CuAlO$_2$ located an indirect band gap at 1.6-1.8\,eV, \emph{i.e.} far below the optical band gap. It was proposed recently that these results might arise from deep defect levels in the band gap from Cu$_{Al}$ antisites\cite{scanlon-defects}. In contrast, newer studies\cite{gilliland-2007} on thicker films propose a value of around 3\,eV for the indirect transition.

Figure~\ref{fig:CuAlO2-sXLDA} (a) shows the band structure of CuAlO$_2$ from our sX-LDA 
calculations in comparison to those from GGA. 
The Cu atoms introduce shallow 3d electrons, which are energetically degenerate with
the weakly dispersive states from non-bonding O 2p orbitals and dominate the upper valence band.
Interaction of Cu 3d and O states locally
push up a mixed Cu-O state above the undispersed background and form a mesa between
the valence band maximum at the F point and a energetically slightly lower
extremum at L. The lower effective mass of this state supports the
formation of shallow acceptor levels. The lowest conduction band shows a marked dispersion with local minima of the conduction band at or near all high-symmetry points. 
GGA, as well as the hybrid functionals B3LYP\cite{robertson-2001} and HSE\cite{vidal-2010,trani-2010,scanlon-defects} and also G$_0$W$_0$\cite{vidal-2010,trani-2010}, predict the global conduction band minimum to be at the $\Gamma$ point, with a rivaling minimum at the $L$ point. A previous study using the sX-functional yielded the same result\cite{xiong-2006}, but underestimated the band gap due to using to strong screening of the Hartree-Fock exchange.

The situation is clearly different in our sX-LDA band structure. The renormalization at the high-symmetry points is remarkably dependent on the contribution of different states to the local band structure. The inclusion of non-local exchange affected particularly the lowest conduction band at the $\Gamma$ and the Z point, which are dominated by Cu $d$ and $s$ states with contributions of O $p$ (Z) and Al $p$ ($\Gamma$). In contrast, the conduction band at the F and the L point consists of Cu $p$ states mixed with Al $s$ (L) and O $p$ (F) and these states experienced only a small renormalization of ~0.3\,eV. As a consequence, the minimum at the $\Gamma$ point is pushed above the minima at $L$ and $F$ points, the global conduction band minimum moves to the $L$ point. We find a minimum indirect band gap of 2.8\,eV between the F and the L point and a slightly larger direct band gap of 2.95\,eV at the L point. The qualitative prediction of the lowest conduction band in our case is very similar to the band structure from self-consistent quasiparticle G$_0$W$_0$@scCOHSEX calculations, as reported recently by Trani \emph{et al}\cite{trani-2010}. The reason might be the similar description of statically screened exchange interaction in sX-LDA and COHSEX, which is different to the (screened) exact exchange portions in HSE and B3LYP. Table~\ref{tab:CuAlO2-gaps} summarizes the indirect and direct band gap as obtained by the different methods.

The valence band draws a similar picture to the conduction band. The lifting of self-interaction causes a strong downward push of the Cu $d$ and O $p$ dominated high-mass bands at the valence band top and leaves a high mesa between the valence band maximum and the L point. This opens the direct band gap at the $\Gamma$-point to a value of 6.8\,eV. The valence Cu $d$ states give rise to a high peak in the partial density of states (PDOS), Fig.~\ref{fig:CuAlO2-sXLDA} (b), which is shifted down by 1.8\,eV compared to GGA. We find the peak maximum at about 3.1\,eV and an additional shoulder at about 2.9-3\,eV below the VBM. Interestingly, this agrees well with recent x-ray photoelectron spectroscopy XPS experiments\cite{aston-2005,scanlon-2009}, where the peak maximum was found at an energy of -2.8 to -3\,eV. The good prediction of the depth of $d$ levels from sX-LDA was previously shown for ZnO and other transition-metal semiconductors\cite{lee-2008,clark-ZnO}. 

A characteristic effect of non-local exchange in isolated atoms is the species-dependent renormalization of energy levels from the lifted self-interaction. To a certain extend, this carries over to solids and manifests in a change of electron negativity difference, the lowering in energy levels and the broadening of the energy spectra. For CuAlO$_2$, we observe a transfer of charge from the aluminium ions to the oxygen cations, which leads to a slight increase in bond polarization and might be responsible for the opening of the band gap. 
The PDOS suggests a strong energetic downshift and broadening of oxygen and aluminium states. 
The oxygen 2$p$ states are shifted down by 2\,eV compared to GGA and broadened down to an energy of ~11\,eV below the valence band maximum. This agrees well to available experimental XPS data\cite{aston-2005}.

We note that HSE and G$_0$W$_0$ apparently place the flat Cu $d$ bands at higher energies compared to sX-LDA, ~1.5-2\,eV below the valence band maximum. This is not surprising in case of G$_0$W$_0$, as this uses perturbation of a LDA ground state and as such acts particularly on the excited states\cite{lee-2007}. Consequently, the under-binding of $d$ states from LDA is maintained. In contrast, hybrid functionals make use of error cancellation by combining LDA/GGA and Hartree Fock exchange and improve on the description of electronic and structural properties. The idea in sX-LDA is to compensate the over-localization of Hartree-Fock by only including a short-range exact exchange contribution. The exponential decay of this contribution accounts for the screening of the electron-electron interaction by the surrounding electron cloud. At the other hand, the self-interaction is removed almost complete near the electron, where the electron correlation is low. We believe that this is the origin of the improved description of $d$ electrons compared to GGA and Hartree-Fock. 

The approach in HSE is at the other end of the scale. Here, the attempt is to counter the unnatural delocalization of PBE by including a portion of Hartree-Fock exchange, usually 25\%, to the short range electron-electron interaction. Hence, the electron self-interaction is only partly lifted at all distances, which might lead to the under-binding of $d$-electrons in this case. At the other hand, this explains the qualitative similarity of the HSE band structures with those from GGA and LDA.

\begin{table}[tb]
\centering
\caption{\label{tab:CuAlO2-gaps} Calculated bandgaps in CuAlO$_2$ from GGA,
various hybrid functionals and quasiparticle calculations. scG$_0$W$_0$(+P) refers to G$_0$W$_0$@scCOHSEX (calculations with model polaron correction). The minimum band gap is indirect in all
methods.}
\begin{tabular*}{\columnwidth}{@{\extracolsep{\fill}} | c | c  c | }
\hline
Method&E$_{ind}$&E$_{dir}$ (L point)\\
\hline
PW91&1.9\, eV&2.6\,eV\\
B3LYP\cite{robertson-2001}&3.9\,eV&4.5 eV\\
HSE06\cite{trani-2010}&3.6\,eV&4.1 eV\\
sX-LDA&2.8\,eV&2.95\,eV\\
G$_0$W$_0$\cite{trani-2010}&3.1\,eV&3.4\, eV\\
scG$_0$W$_0$\cite{trani-2010}&5.0\,eV&5.1\,eV\\
scG$_0$W$_0$+P\cite{trani-2010}&3.8\,eV&3.9\,eV\\
Exp&3.0\,eV&3.5\,eV\\
\hline
\end{tabular*}
\end{table}

\begin{figure}
\begin{minipage}{0.499\columnwidth}
\centering
\includegraphics*[width=\columnwidth]{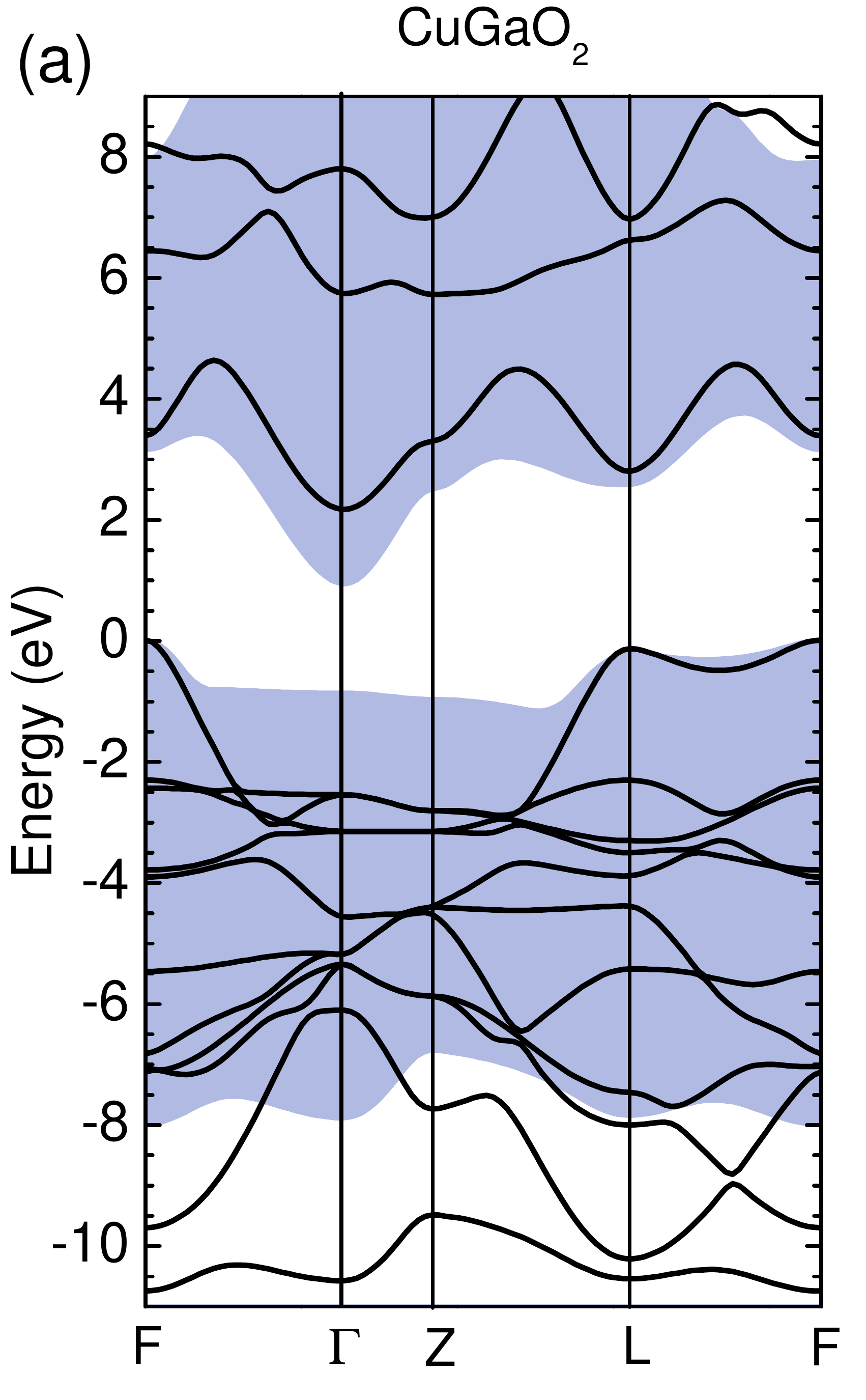}
\end{minipage}
\space
\begin{minipage}{0.477\columnwidth}
\centering
\includegraphics*[width=\columnwidth]{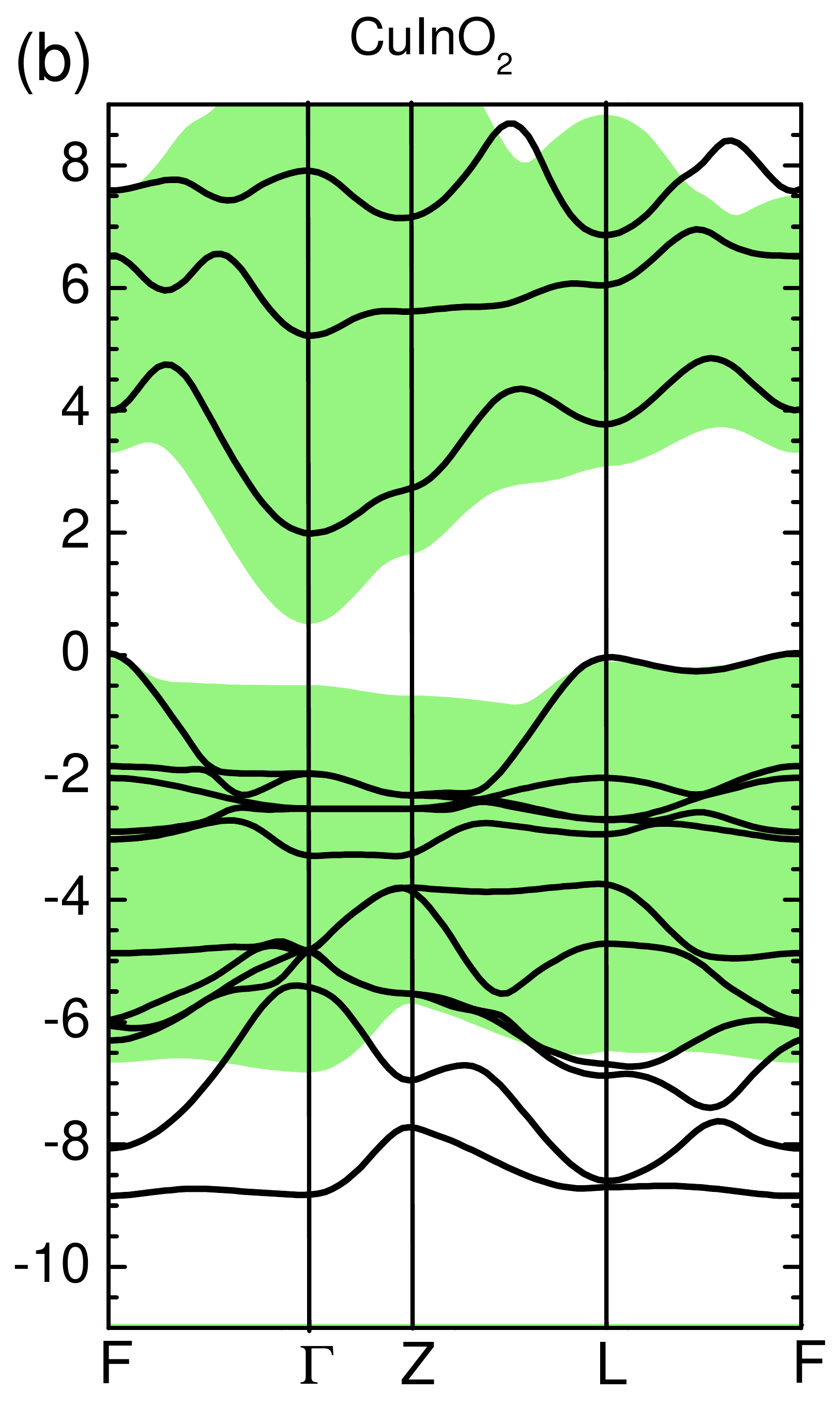}
\end{minipage}
\caption{\label{fig:CuGaInO2-bands} (Color online) Comparison of sX-LDA band structures of (a) CuGaO$_2$ and (b) CuInO$_2$. The shaded areas represent the valence and conduction bands from GGA calculations.}
\end{figure}

\begin{figure}
\begin{minipage}{0.48\columnwidth}
\centering
\includegraphics*[width=\columnwidth]{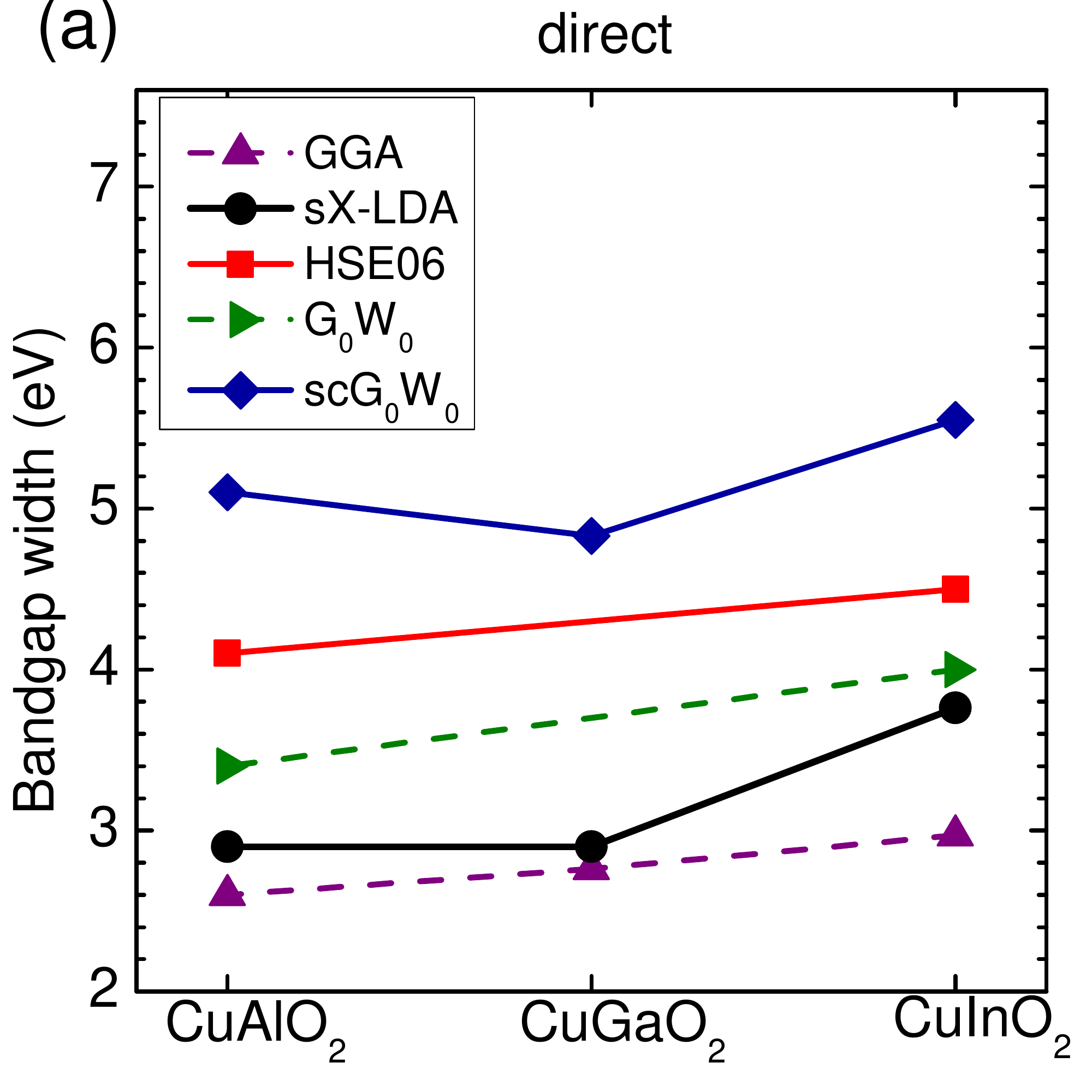}
\end{minipage}
\space
\begin{minipage}{0.48\columnwidth}
\centering
\includegraphics*[width=\columnwidth]{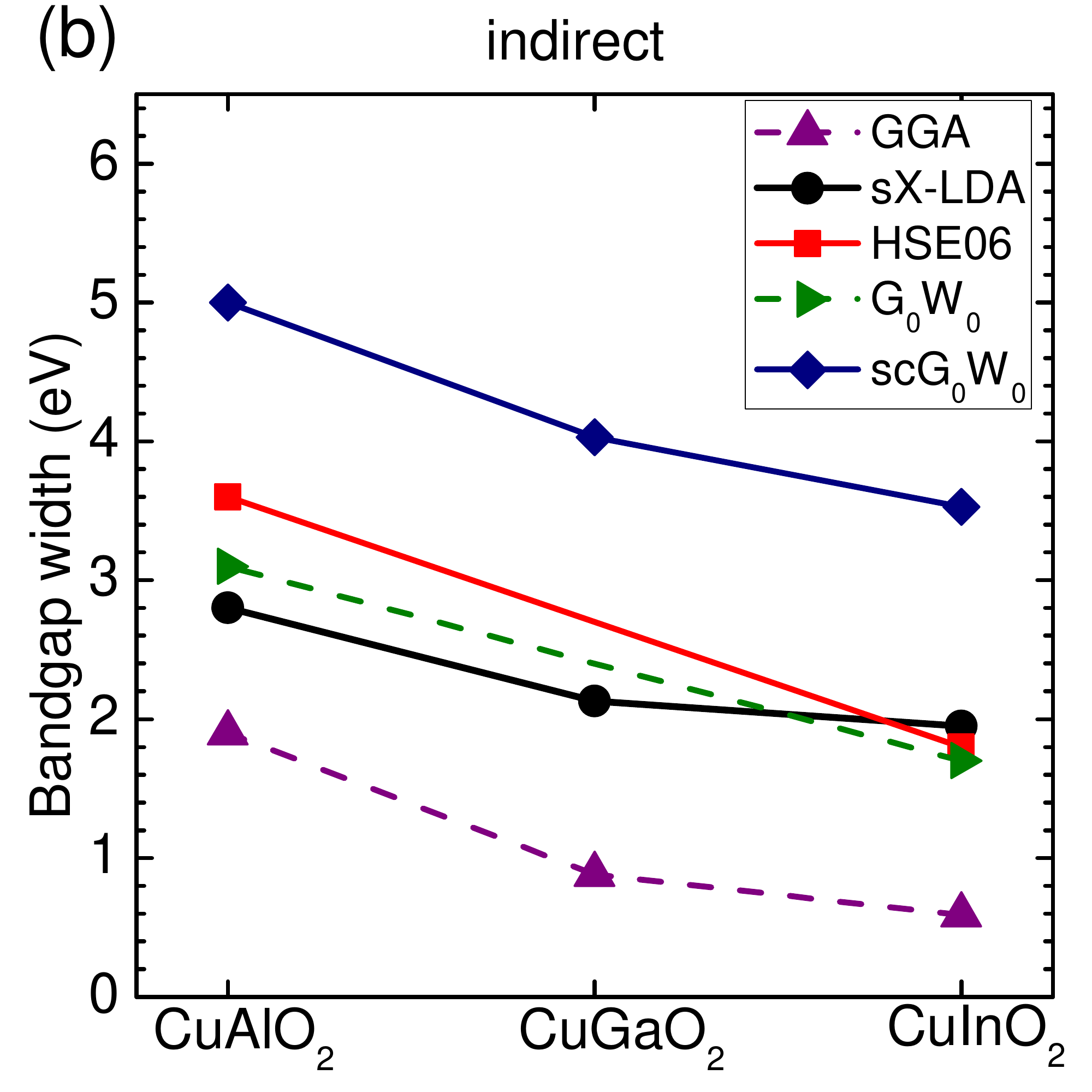}
\end{minipage}
\caption{\label{fig:trends} (Color online) Trends of fundamental (a) direct and (b) indirect band gaps of CuMO$_2$ (M=Al,Ga,In) with the atomic mass of M. For brevity, scG$_0$W$_0$ refers to G$_0$W$_0$@scCOHSEX.}
\end{figure}

\begin{table}
\centering
\caption{\label{tab:VBM-trends} Valence band widths as obtained from GGA, sX-LDA and experiments.}
\begin{tabular*}{\columnwidth}{@{\extracolsep{\fill}} | c | c c c| }
\hline
Approximation&CuAlO$_2$&CuGaO$_2$&CuInO$_2$\\
\hline
GGA&~8\,eV&~8\,eV&~7\,eV\\
sX-LDA&~10.75\,eV&~10.7\,eV&~9\,eV\\
Exp\cite{shin-2009}&~11\,eV&~10\,eV&9.5\,eV\\
\hline
\end{tabular*}
\end{table}

In the optical spectra, replacing Al by heavier group III atoms like Ga or In seems to shift the onset of optical absorption to higher energies. The reported optical band gaps for CuGaO$_2$ and CuInO$_2$ are 3.6\,eV\cite{ueda-2001} and 3.9\,eV\cite{yanagi-2001}, respectively, whereas the direct gap in CuAlO$_2$ is controversial but likely around 3.5\,eV. At first glance, this seems to contradict the trend in other oxides, where the optical band gap decreases with the atomic number. Based on LAPW-LDA calculations, Nie \emph{et al.}\cite{nie-2002} suggested that the observed absorption onset corresponds not to the minimum band gaps but to direct optical transitions at the L point, as the direct transitions at $\Gamma$ and $Z$ point are symmetry-forbidden. Indeed, GGA band structures for CuGaO$_2$ and CuInO$_2$ 
show that the conduction band minimum at the $\Gamma$ point moves towards lower energies for increasing atomic number. This is due to the contribution of antibonding $s$ states at these points, which become energetically more favorable as the volume of the unit cell increases.
At the other hand, the local conduction band minimum near the L point smoothens and becomes less prominent, thus effectively increasing the optically active direct gap at the L point. Hybrid functional and GW calculations exhibit a noticeable change in dispersion of the lowest conduction band compared to LDA/GGA, renormalizing the energies at the $\Gamma$ point stronger than at the L point. As a result, the minimum direct band gaps are now located at the L point. We find the same behavior in our sX-LDA calculations, see Fig.~\ref{fig:CuGaInO2-bands} (a) and (b). 
While the different renormalization of points with and without $d$-contributions narrows the lowest conduction band, the minima at L and F point become more prominent and preferable points for optical transitions. 

\begin{figure}
\begin{minipage}{0.488\columnwidth}
\centering
\includegraphics*[width=\columnwidth]{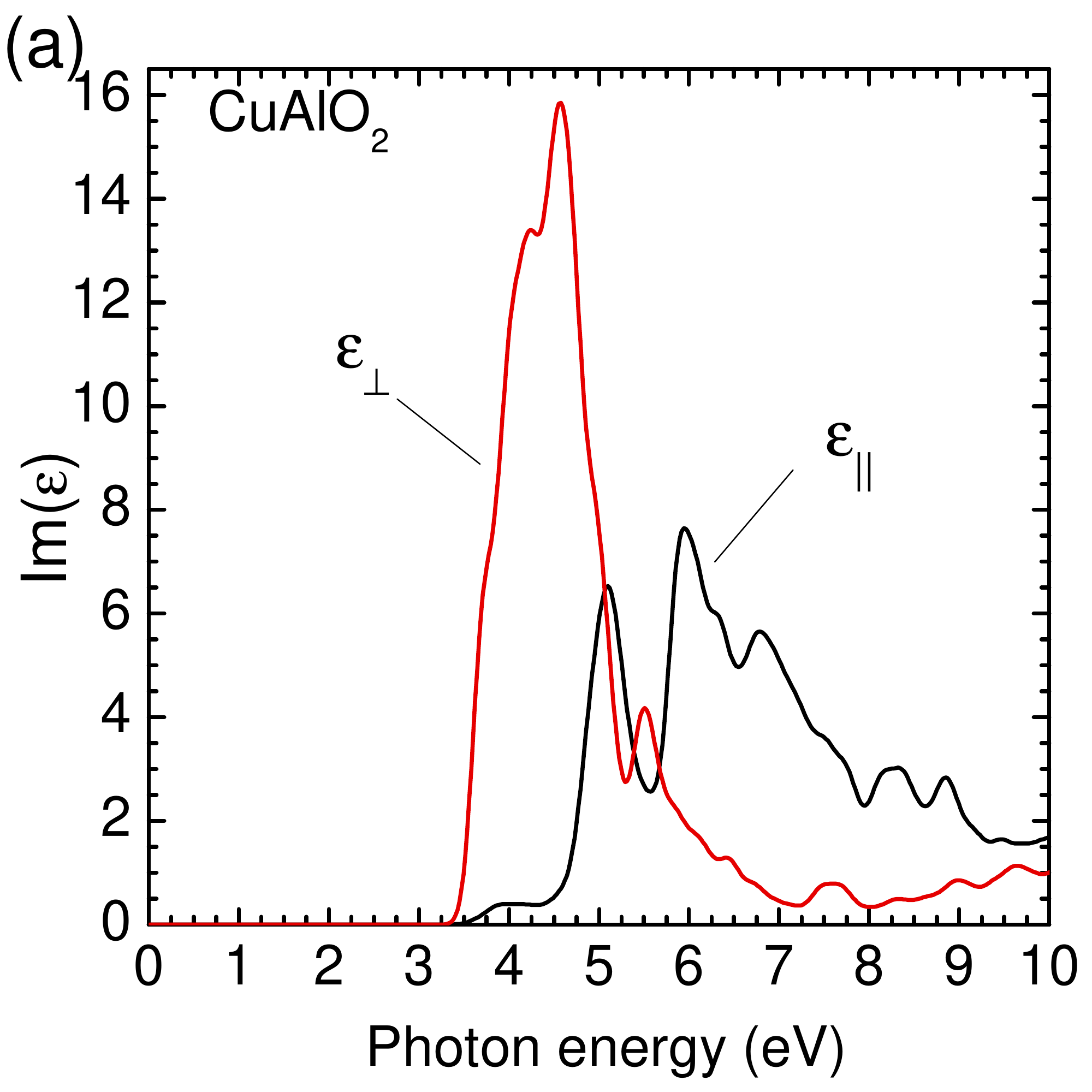}
\end{minipage}
\space
\begin{minipage}{0.488\columnwidth}
\centering
\includegraphics*[width=\columnwidth]{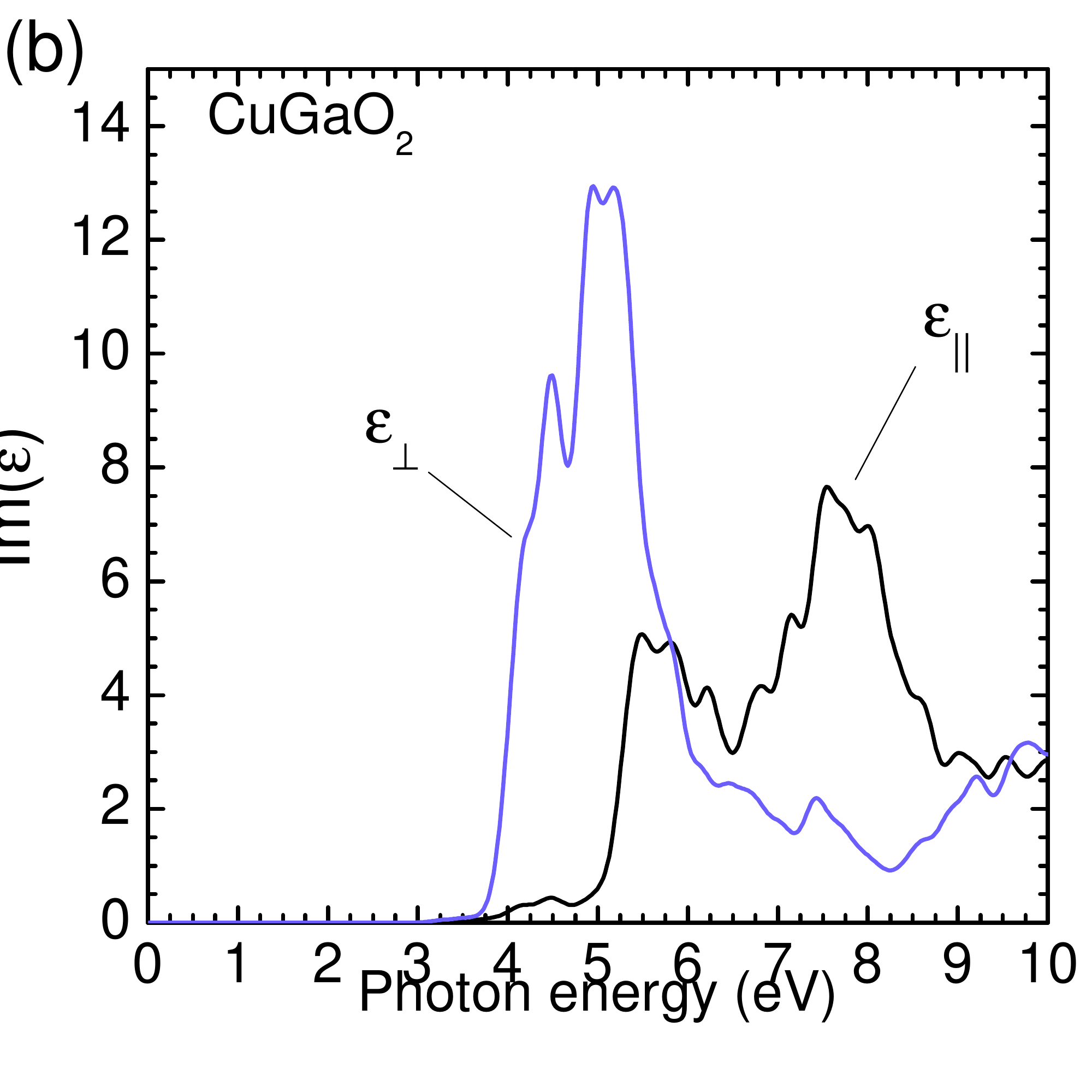}
\end{minipage}
\\
\begin{minipage}{0.488\columnwidth}
\centering
\includegraphics*[width=\columnwidth]{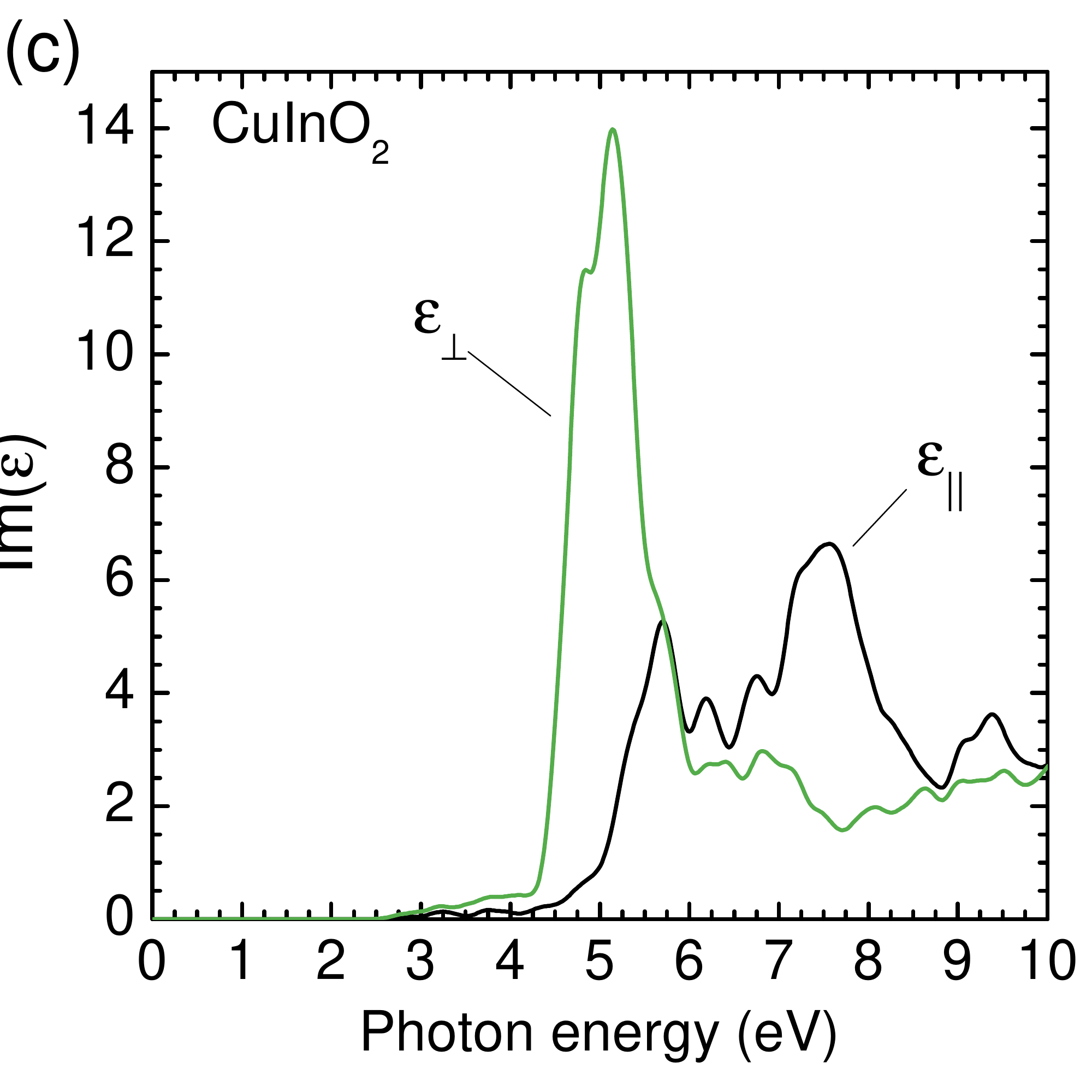}
\end{minipage}
\space
\begin{minipage}{0.488\columnwidth}
\centering
\includegraphics*[width=\columnwidth]{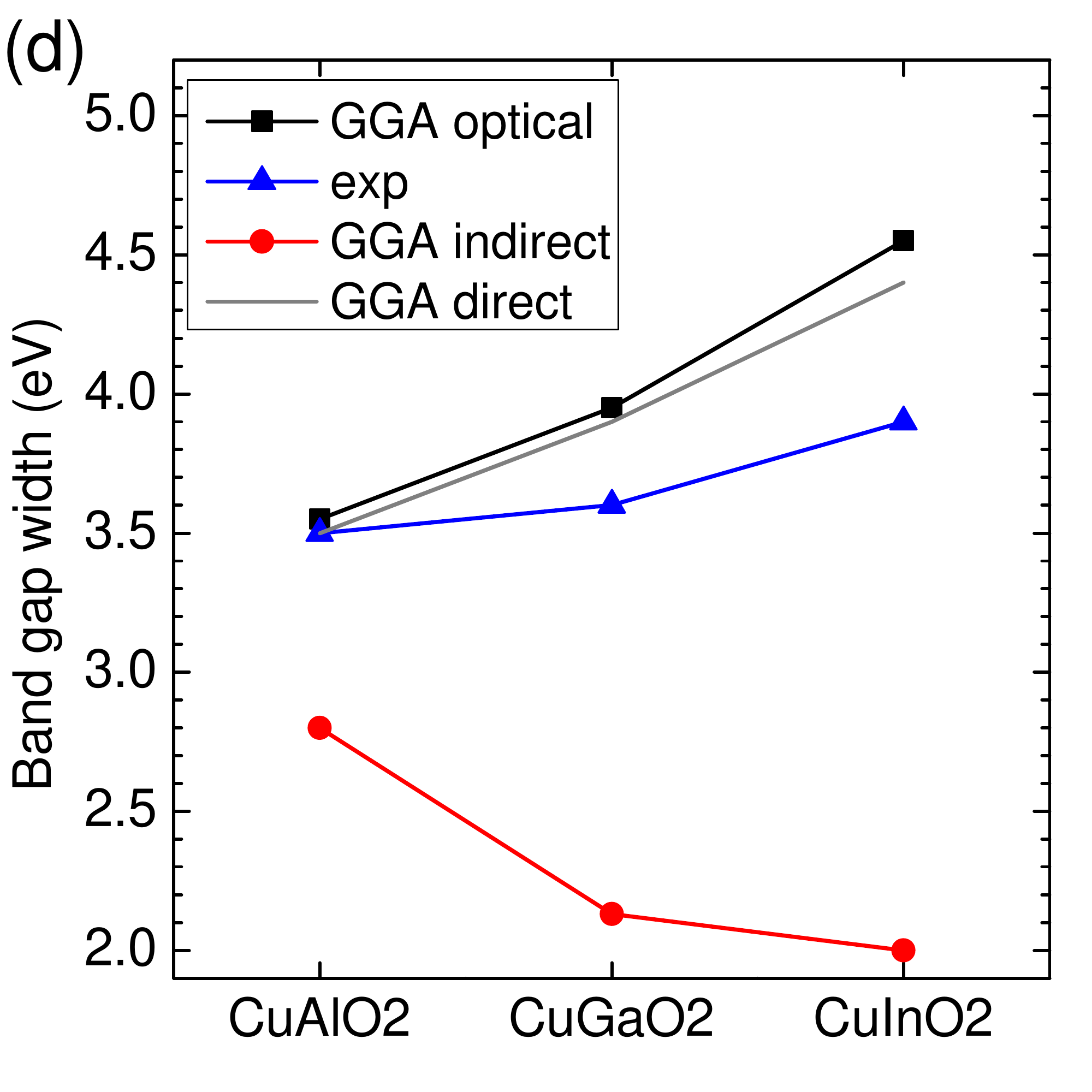}
\end{minipage}
\caption{\label{fig:CuAlGaInO2-epsilon} (Color online) Imaginary part of the dielectric function for light polarizated parallel ($\epsilon_{\parallel}$) and perpendicular ($\epsilon_{\perp}$) to the $c$-axis from GGA calculations on (a) CuAlO$_2$, (b) CuGaO$_2$ and (c) CuInO$_2$. The peaks were broadened by a Gaussian smearing of 0.1 eV and the minimum indirect gaps were adjusted by a scissor operator to match the sX-LDA values. (d) Comparison of the optical band gaps (squares, obtained by a ($\alpha h\nu$)-$h\nu$ plot of the absorption spectra corresponding to (a)-(c) with experimental values (triangles) and minimum indirect band gaps (circles).}
\end{figure}

Fig.~\ref{fig:trends} compares the calculated direct and indirect band gaps at the L point from different methods. Our calculated band gap for CuInO$_2$ of 3.8\,eV compares nicely with the experimentally measured optical band gap of 3.9\,eV, only G$_0$W$_0$ being closer. HSE06 and G$_0$W$_0$@scCOHSEX overestimate the band gap, while GGA, as expected, underestimates. 
The available results for CuGaO$_2$ are less favorable. Trani \emph{et al.}\cite{trani-2010} reported that G$_0$W$_0$@scCOHSEX gives a band gap size smaller than the one in CuAlO$_2$, thus breaking the tendency of the predicted optical band gaps to increase with the atomic mass. 
Similarly, the value from our sX-LDA calculations is almost identical to the band gap in CuAlO$_2$, when it should be between the values of CuAlO$_2$ and CuInO$_2$. For sX-LDA, this also shows in the trends of the minimum indirect band gaps as illustrated in Fig.~\ref{fig:trends} (b). 
Our calculated sX-LDA band gap for CuGaO$_2$, 2.13\,eV, is near the value for CuInO$_2$, 2.0\,eV. The applied screening in our calculation obviously is slightly too strong to reproduce the experimental trends.
Unfortunately, we found no data for HSE or G$_0$W$_0$ values of the band gaps in CuGaO$_2$.
Overall, all methods predict a noticeable decrease in indirect band gap size with atomic weight. Sasaki \emph{et al.}\cite{sasaki-2003} found in CuInO$_2$ an indirect band of 1.44\,eV, in sight of the predicted band gaps from G$_0$W$_0$, sX-LDA, HSE and even B3LYP. As suggested for CuAlO$_2$, deep defect levels or excitons might lead to a shift of the measured indirect band gap towards lower energies.\\
Similar to the indirect band gap, the valence band width also exhibits a decreasing trend in the sequence CuAlO$_2$>CuGaO$_2$>CuInO$_2$, which roots in the increase in Cu-O bond lengths\cite{shin-2009}. GGA reproduces the trend but compresses the valence bands of all three materials by several eV, see Table~\ref{tab:VBM-trends}.  Our sX-LDA calculations show an appreciable improvement in the absolute values, which are close to the experimentally obtained band widths in the range 9-11\,eV.

Finally, we have performed optical calculations to obtain the imaginary part of the dielectric function $\epsilon$ of the three systems. Unfortunately, optical calculations employing sX-LDA are prohibitively expensive and we had to restrict our investigations to GGA properties. As a compromise, we scaled the minimum band gaps from GGA to the sX-LDA values. Figure~\ref{fig:CuAlGaInO2-epsilon} shows the obtained imaginary part of the dielectric function for light polarized parallel and perpendicularly to the $c$-axis of the crystal. In accordance to the absorption spectra from 
 Nie \emph{et al}\cite{nie-2002}, we observe a noticeable anisotropy with respect to the light polarization. 
As the absorption coefficient is proportional to Im($\epsilon$), significant absorption of light polarization along the layer stacking is retarded towards higher energies, with a tail reaching down to the optical band gap energy. Similarly, the onset of appreciable absorption is located at an energy of 3.5-4.5\,eV for all materials, confirming that transitions involving the $\Gamma$ and $Z$ points provide no or only a small contribution to the low energy absorption. 
We have obtained the optical band gap energies by means of Tauc plots\cite{tauc-1968} of the corresponding absorption spectra. For direct transitions, this yields optical band gaps of 3.55\,eV, 3.9\,eV and 4.5\,eV for CuAlO$_2$,CuAlO$_2$ and CuAlO$_2$, respectively. 
These values fit well to the direct band gaps at L and F points. Figure~\ref{fig:CuAlGaInO2-epsilon} (d) compares the obtained optical band gaps with experimental values and the direct band gaps at the L point. The agreement is very well for CuAlO$_2$, but our corrected GGA overestimates the optical band gaps of CuGaO$_2$ and CuInO$_2$ by 0.3\,eV and 0.5\,eV, respectively. 

\subsection{Band structure of CuCrO$_2$}
\begin{figure}
\begin{minipage}{0.488\columnwidth}
\centering
\includegraphics*[width=\columnwidth]{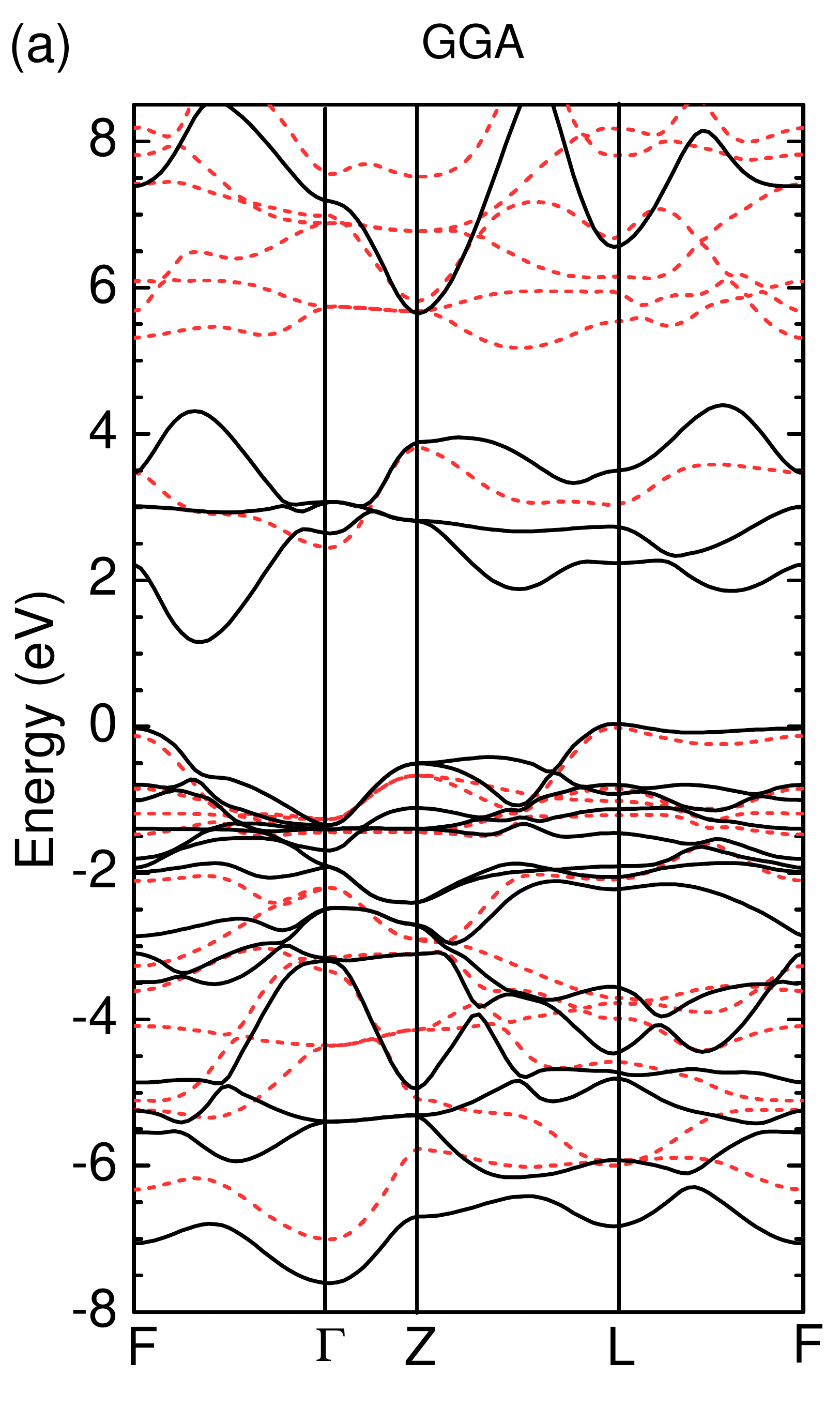}
\end{minipage}
\space
\begin{minipage}{0.488\columnwidth}
\centering
\includegraphics*[width=\columnwidth]{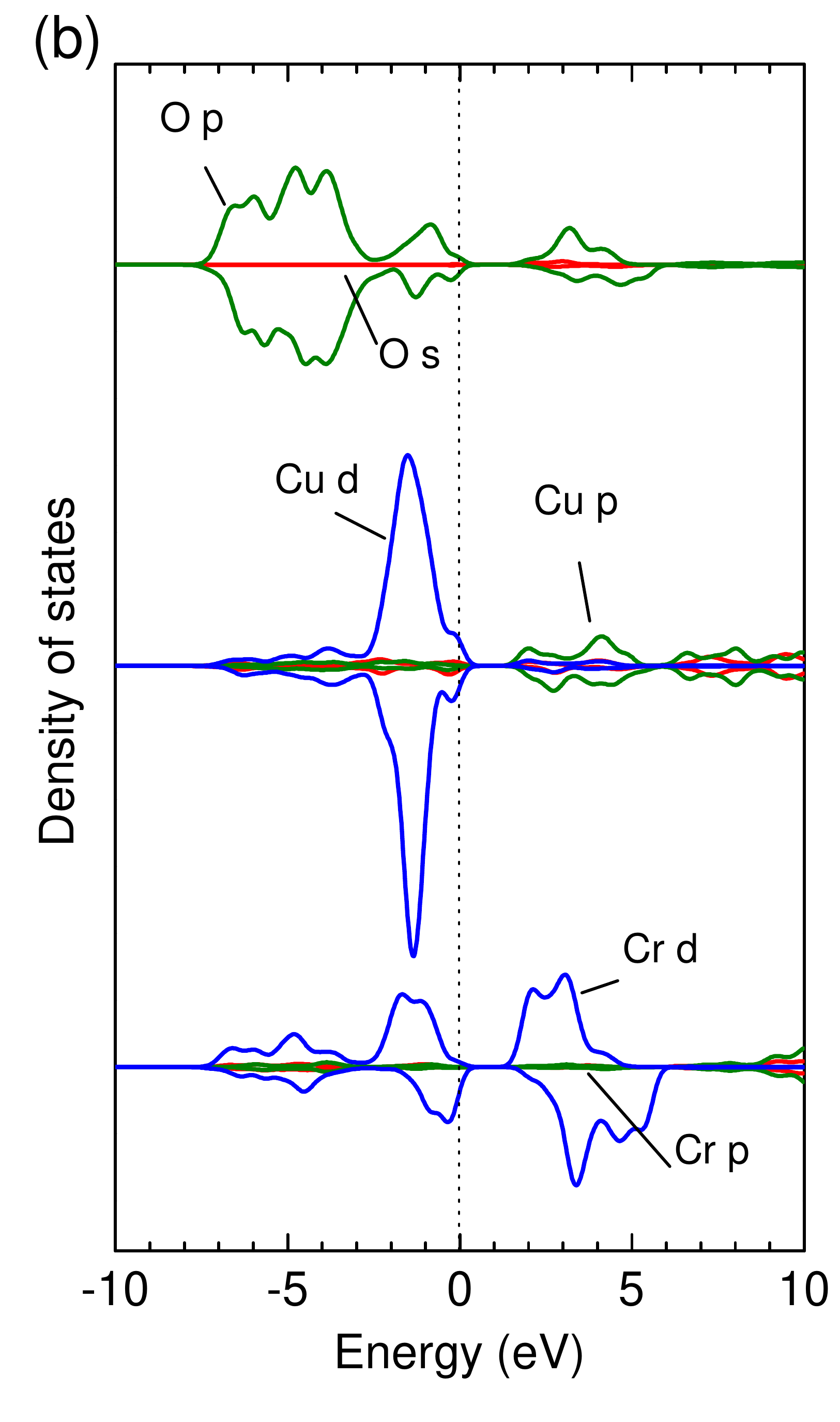}
\end{minipage}
\caption{\label{fig:bands-CuCrO2} (Color online) (a) Electronic band structure and (b) corresponding PDOS of CuCrO$_2$ from GGA calculations. The black solid lines and the (red) broken lines represent states of up-spin and down-spin direction electrons, respectively.}
\end{figure}

\begin{figure}
\begin{minipage}{0.488\columnwidth}
\centering
\includegraphics*[width=\columnwidth]{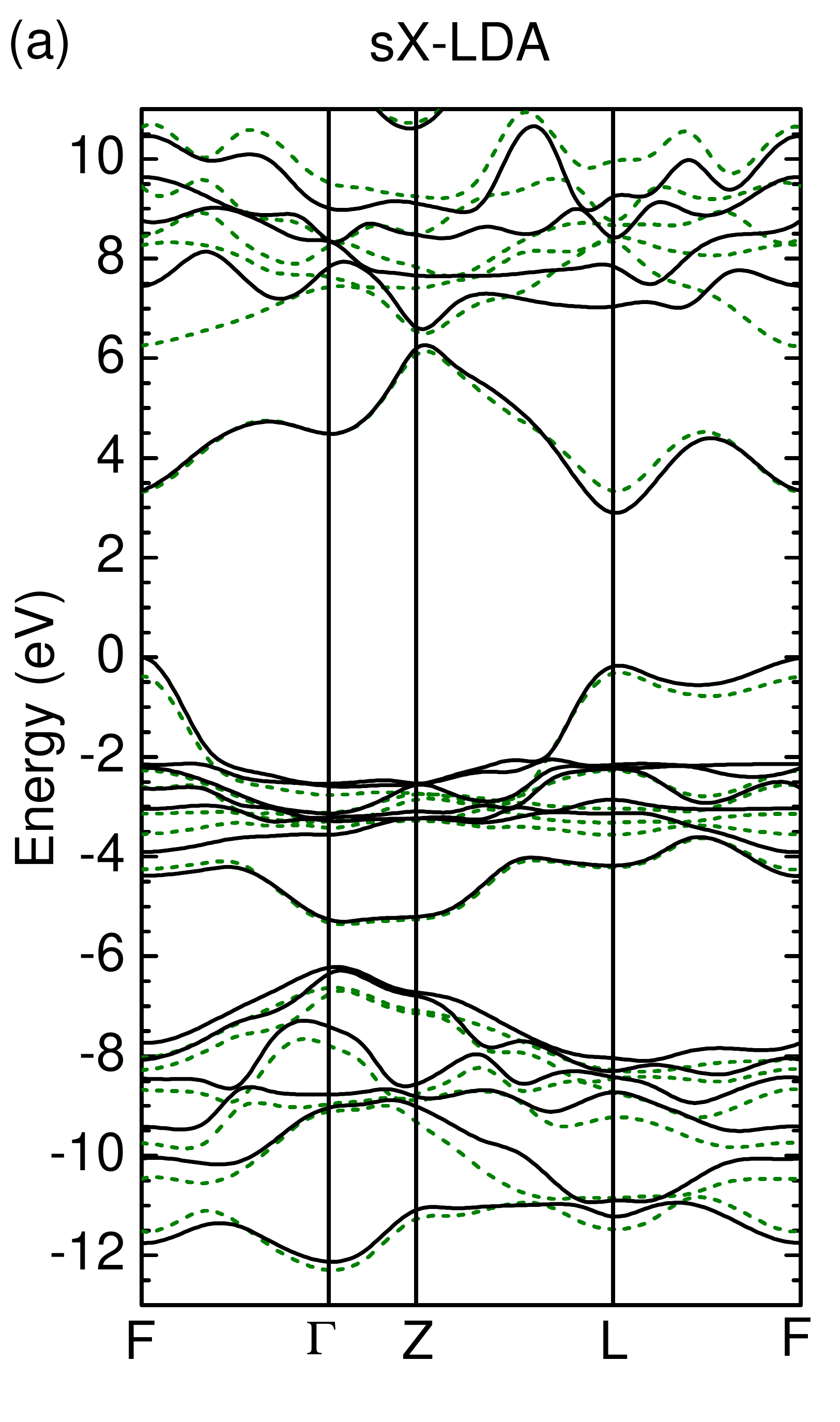}
\end{minipage}
\space
\begin{minipage}{0.488\columnwidth}
\centering
\includegraphics*[width=\columnwidth]{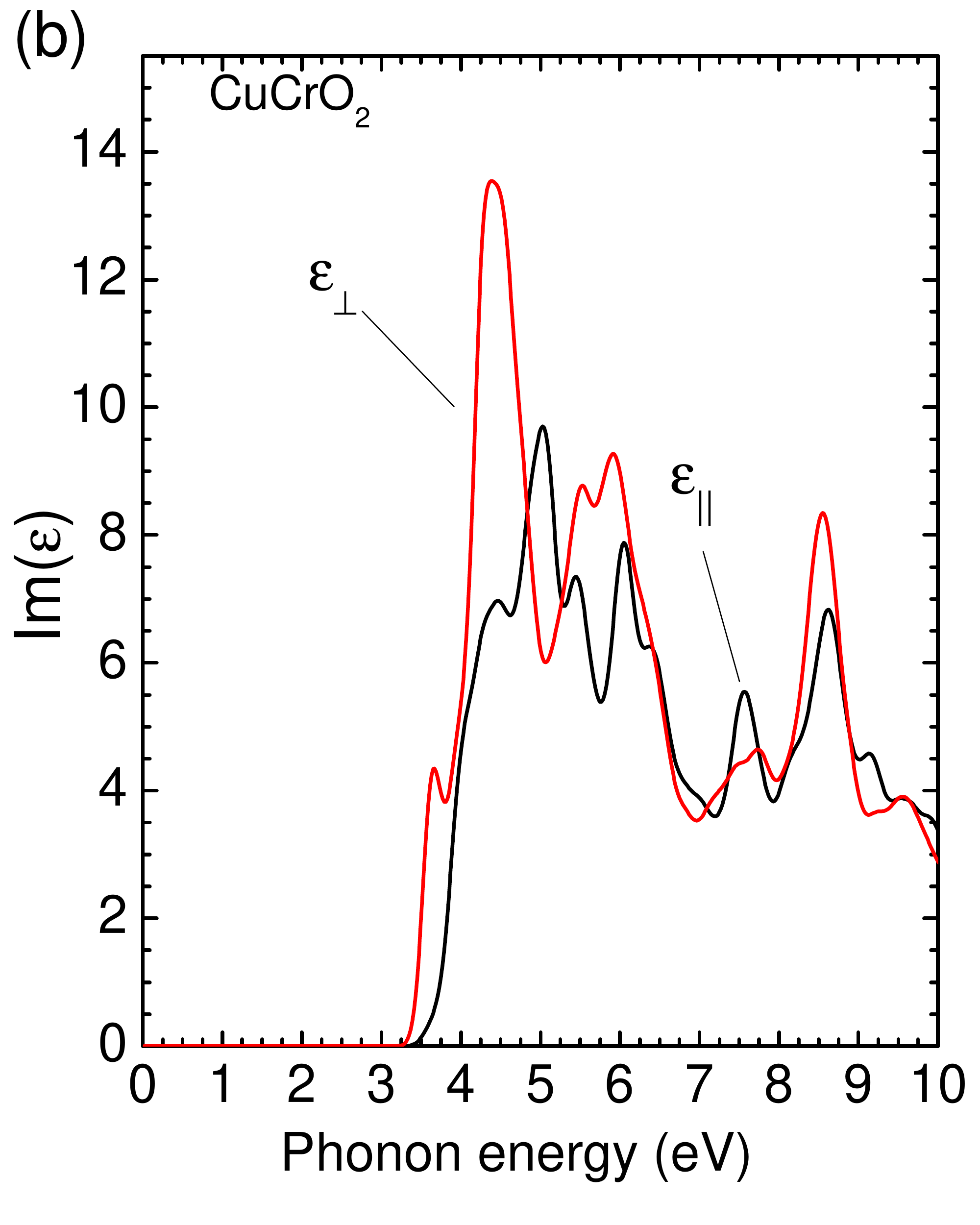}
\end{minipage}
\caption{\label{fig:CuCrO2-pdos-epsilon} (Color online) (a) sX-LDA band structure of CuCrO$_2$. As in Fig.~\ref{fig:bands-CuCrO2}, the black solid lines and the (red) broken lines represent states of up-spin and down-spin direction electrons, respectively. (b) Imaginary part of the dielectric function and corresponding absorption spectrum from GGA calculations. Figure description as in Fig.~\ref{fig:CuAlGaInO2-epsilon}.}
\end{figure}

Another group of Cu-based transparent conducting oxides contains particular transition metals,  \textit{e.g.} Fe or Cr, instead of elements from the third group. The electrons in the partly filled $d$-orbitals might then form a magnetic order and thus allow for transparent materials with magnetic properties. Aside from magnetic properties, Mg-doped CuCrO$_2$ possesses the highest conductivity of all reported delafossites\cite{tate-2002}. Studies on CuCrO$_2$ found an absorption onset energy of about 3\,eV\cite{benko-CuCrO2,nagarajan-2001}, which renders CuCrO$_2$ to be transparent, and additional indirect transitions at 1.28\,eV\cite{benko-CuCrO2} and 1.45\,eV\cite{zheng-2006}. The nature of the optical transition at 3\,eV is not fully clear. Early photoelectrochemical measurements\cite{benko-CuCrO2} suggest an indirect gap, whereas recent absorption measurements point toward a direct transition. 

We have calculated the ground state and the electronic band structure for CuCrO$_2$ by means of GGA and sX-LDA. The Cr atoms are spin-polarized and impose an anti-ferromagnetic order on the ground state of the crystal, which we confirmed on a 2x2x2 rhombohedral supercell. The energy difference between anti-ferromagnetic and ferromagnetic spin order, however, was found to be of order 1\,meV, so chose to model CuCrO$_2$ by the rhombohedral unit cell in our calculations to save computational ressources.
Figure~\ref{fig:bands-CuCrO2} (a) shows the band structure of CuCrO$_2$ from GGA calculations. 
Due to the ferromagnetic ground state, three single electron Cr $d$ states of $\alpha$-spin appear close to the valence band top and distort local the mixture of Cu $d$ and O $p$ bands. The bottom of the conduction band includes the remaining two $\alpha$-spin Cu $d$ states. The complimentary $\beta$-spin bands are shifted deep in the conduction band due to the exchange splitting. This also shows in the PDOS in Fig.~\ref{fig:bands-CuCrO2} (b). The global conduction band minimum of 1.12\,eV is moved between the F and the $\Gamma$ point, the fundamental band gap being indirect and between this point (X point) and the F point. The minimum direct gap of 1.97\,eV is at the X point and differs from the indirect band gap by the difference in the valence band top. The qualitative positions of the fundamental transitions are in good agreement with the band structures of Scanlon \emph{et al}\cite{scanlon-2009}, who used a hexagonal 2H unit cell and GGA+U to simulate CuCrO$_2$ and obtained an anti-ferromagnetic ground state.

Screened exact exchange noticeably changes the predicted band structure, see Fig.~\ref{fig:CuCrO2-pdos-epsilon} (a). In general, the respective band structures of $\alpha$- and $\beta$-spin electrons become quite similar, in contrast to the GGA. As for the other CuMO$_2$, the energies at the $d$-state dominated $\Gamma$ and Z points are strongly shifted up and the global minimum moves to the L point, with a rivaling distinctive minimum appearing at the $F$ point. As a consequence, the indirect band gap opens to 2.9\,eV and is between $F$ and $L$ point, the minimum direct band gap is 3.1\,eV and at the $L$ point. Another direct band gap occurs at the $F$ and is 3.25\,eV wide. Scanlon \emph{et al.}\cite{scanlon-2009} report a value of 2.04\,eV for the fundamental (indirect) gap, but find a larger difference between indirect and direct gap due to the comparatively weak pushdown of the Cu $d$ states in their calculations. Table~\ref{tab:CuCrO2-gaps} summarizes the obtained band gap sizes. As a downside, the agreement of the valence band width with experiment in sX-LDA is worse than in GGA. Arnold \emph{et al.}\cite{arnold-2009} have reported a value of ~8.2\,eV for the valence band width of CuCrO$_2$, which is a considerably lower value than would be expected from the Cu-O bond length. GGA captures this value quite well, whereas sX-LDA give a width of 12\,eV.

Finally, Fig.~\ref{fig:CuCrO2-pdos-epsilon} (b) shows the imaginary part of the dielectric function for light polarized perpendicularly and parallel to the $c$-axis. The anisotropy as observed for the other systems is present but less pronounced. This indicates a fairly homogenous polarizability of CuCrO$_2$ within and perpendicular to the Cr-O layers.

\begin{table}[tb]
\centering
\caption{\label{tab:CuCrO2-gaps} Minimum indirect and direct band gaps of CuCrO$_2$ as obtained from GGA, GGA+U and sX-LDA calculations and experiments.}
\begin{tabular*}{0.8\columnwidth}{@{\extracolsep{\fill}} | c | c  c | }
\hline
&E$_{ind}$&E$_{dir}$ (L point)\\
\hline
GGA&1.12\,eV&1.97\,eV\\
GGA+U\cite{scanlon-2009}&2.04\,eV&2.55\,eV\\
sX-LDA&2.9\,eV&3.1\,eV\\
Exp&1.28\,eV\cite{benko-CuCrO2}, 1.45\,eV\cite{zheng-2006}&3.1\,eV\cite{benko-CuCrO2}\\
& 3.08\,eV\cite{benko-CuCrO2}&\\
\hline
\end{tabular*}
\end{table}

\section{Conclusion}
We found a remarkable improvement of the electronic properties of CuMO$_2$ (M=Al,Ga,In,Cr) from the use of the screened exchange functional. The electronic band gaps, valence band widths and the binding levels of localized electrons (Cu $d$ and O $p$) are in good agreement with experiments.

\section{Acknowledgements}
This work was supported by funds from the EU project 'Orama'.

\bibliographystyle{apsrev4-1}

\end{document}